\documentstyle[12pt]{article}

\renewcommand{\epsilon}{\varepsilon}
\input{tcilatex}

\begin{document}

\title{Probability and Nonlocality in Many Minds Interpretations of Quantum
Mechanics}
\author{Meir Hemmo \and Itamar Pitowsky}
\date{}
\maketitle

\bigskip

\begin{center}
ABSTRACT
\end{center}

\noindent We argue that a certain type of many minds (and many worlds)
interpretations of quantum mechanics, e.\,g.\ Lockwood ([1996a]), Deutsch
([1985]) do not provide a coherent interpretation of the quantum mechanical
probabilistic algorithm. By contrast, in Albert and Loewer's ([1988])
version of the many minds interpretation there is a coherent interpretation
of the quantum mechanical probabilities. We consider Albert and Loewer's
probability interpretation in the context of Bell-type and GHZ-type states
and argue that it implies a certain (weak) form of nonlocality.

\section{Introduction}

\begin{center}
\label{intro}
\end{center}

In this paper we shall consider two questions in the context of many minds
interpretations of quantum mechanics. The first question is whether and how
the notion of probability makes sense in these interpretations. We shall
mainly refer to two versions of many minds interpretations: Albert and
Loewer's ([1988]) stochastic version in which the minds don't supervene on
physical states, and Lockwood's ([1996a,b]) version in which there is full
supervenience. These two versions have been discussed in some detail in a
special symposium hosted by this Journal ([1996], Vol. 47, pp. 159-248).
Lockwood's approach to probabilities seems to be accepted amongst many
authors in the many worlds tradition (though with different styles) e.\,g.\
Deutsch ([1985]), Zurek ([1993]), Saunders ([1998]), Papineau ([1996]),
Vaidman ([1998]), and others. Our second question concerns the implications
of the notion of probability in Albert and Loewer's theory these approaches
on the question of nonlocality. In particular the questions is: what are the
implications of Bell's theorem and the GHZ (Greenberger, Horne and Zeilinger
[1989]) set up in Albert and Loewer's stochastic theory?

The structure of the paper is as follows. We first briefly present Albert
and Loewer's version of the many minds interpretation and we set up the
problem of interpreting the probabilities in a many minds (worlds) picture
(section \ref{al}). In Section \ref{lock} we present and discuss the
supervenience versions (focusing on Lockwood) of the many minds
interpretation, and we argue that in these versions the probability
interpretation is wanting. Then in section \ref{chance} we argue that the
Albert-Loewer many minds interpretation implies a certain weak form of
nonlocal correlations between subsets of minds. Finally, in Section \ref{ghz}
we demonstrate the nonlocality of the Albert-Loewer interpretation using the
Greenberger, Horne and Zeilinger (GHZ) ([1989]) set up.

To get a quick grip on many minds interpretations consider the scheme of a
generic (impulsive) measurement of the $z$-spin variable of an electron in
noncollapsing quantum mechanics. Take a composite of quantum system,
apparatus and observer (respectively) $S+M+O$ initially in the state 
\begin{equation}
|\Psi _{0}\rangle =\Big(\alpha |-_{z}\rangle +\beta |+_{z}\rangle \Big)%
\otimes |\psi _{0}\rangle \otimes |\Phi _{0}\rangle ,  \label{eq:zero}
\end{equation}
where $\left| \alpha \right| ^{2}+\left| \beta \right| ^{2}=1$. Here the $%
|\pm _{z}\rangle $ are the $z$-spin eigenstates, $|\psi _{0}\rangle $ is the
ready state of $M$ and $|\Phi _{0}\rangle $ is some suitable state of $O$'s
brain initiating conscious mental states. We assume that the evolution of
the global state is described by the Schr\"{o}dinger equation alone,
i.\thinspace e.\ there is no collapse of the quantum state. The measurement
interaction between $S$ and $M$ takes the state (\ref{eq:zero}) to the
superposition 
\begin{equation}
|\Psi _{1}\rangle =\Big(\alpha |+_{z}\rangle \otimes |\psi _{+}\rangle
+\beta |-_{z}\rangle \otimes |\psi _{-}\rangle \Big)\otimes |\Phi
_{0}\rangle ,  \label{eq:int1}
\end{equation}
where as can be seen a one-to-one correlation is brought about between the
spin states $|\pm _{z}\rangle $ and the pointer states $|\psi _{\pm }\rangle 
$, but in such a way that the quantum states of both $S$ and $M$ become
entangled in (\ref{eq:int1}). The interaction between $M$ and the observer $%
O $ takes the global state to the final superposition 
\begin{equation}
|\Psi _{f}\rangle =\alpha |+_{z}\rangle \otimes |\psi _{+}\rangle \otimes
|\Phi _{+}\rangle +\beta |-_{z}\rangle \otimes |\psi _{-}\rangle \otimes
|\Phi _{-}\rangle \Big),  \label{eq:int2}
\end{equation}
where the $|\Phi _{\pm }\rangle $ are the observer's {\em brain} states
corresponding to her {\em mental states}.\footnote{%
We shall assume throughout that the set of brain states corresponding to all
possible outcomes of all possible experiments forms a basis in the brain's
Hilbert space. This is a preferred basis in the Hilbert space corresponding
to the (subjective) mental states associated with conscious perception.} As
can be seen the state (\ref{eq:int2}) is now also entangled and the reduced
state of the observer is truly mixed. If one takes this theory to be
complete {\em simpliciter} (called by Albert ([1992]) the bare theory), then
one faces the measurement problem since the measurement has no definite
result. On this view the quantum statistical algorithm which is an algorithm
about the probabilities of measurement results makes no sense (see Albert (%
{\em ibid}, Chapter 6) for more details). Thus in order to avoid the
measurement problem, one needs somehow to supplement the bare theory's
description.

\section{Albert and Loewer's Interpretation}

\begin{center}
\label{al}
\end{center}

Many minds interpretations take the bare theory to be indeed complete and
exactly true but only with respect to the {\em physics,} including the
physics of the brain. In other words, the quantum state never collapses and
no hidden variables are added to the quantum mechanical description. To
supplement the bare theory, further assumptions are made with respect to the
relation between the brain states $|\Phi {\pm }\rangle $ and $O$'s mental
states. Albert and Loewer [1988] make the following two assumptions:

\begin{description}
\item[AL1]  The brain states $\vert \Phi_{\pm}\rangle$ corresponding to $O$%
's mental states are associated at all times with a continuous {\em infinity}
of nonphysical entities called {\em minds} (even for a single observer).

\item[AL2]  Minds do not obey the Schr\"{o}dinger evolution (in particular,
the superposition principle) but evolve in time in a genuinely probabilistic
fashion. For a given measurement, involving a conscious observer, there is
one specific probability measure, given by the Born rule, that prescribes
the chances for each mind to evolve from an initial $\vert \Phi_0\rangle$ to
a final brain state $|\Phi_{i}\rangle$.
\end{description}

In the measurement scheme above each single mind corresponds initially to
the state $|\Phi _{0}\rangle $ and evolves in a {\em stochastic} fashion to
one of the two final brain-mental states $|\Phi {\pm }\rangle $ with the
usual Born probabilities: $\left| \alpha \right| ^{2}$ for a $+$ result and $%
\left|\beta\right|^2$ for a $-$ result. The divergence of the minds occurs
during the evolution of the global state from (\ref{eq:int1}) to (\ref
{eq:int2}). Let us denote by $|\Phi (m)\rangle $ a quantum brain state
indexed by a subset $m$ of the set of minds. The complete description of the
post-measurement state includes the quantum state, and the corresponding
subsets of the set of minds. Therefore, one needs to replace (\ref{eq:int2})
with 
\begin{equation}
|\Psi _{f}(m,n)\rangle =\alpha |+_{z}\rangle \otimes |\psi _{+}\rangle
\otimes |\Phi _{+}(m)\rangle +\beta |-_{z}\rangle \otimes |\psi _{-}\rangle
\otimes |\Phi _{-}(n)\rangle .  \label{eq:flip}
\end{equation}
Here we use the notation $\Psi _{f}(m,n)$, to make explicit the
Albert-Loewer idea that the quantum brain states correspond to, and are
indexed by, subsets of the set of minds. In the state (\ref{eq:flip}) we see
that the minds in the subset $m$ follow the brain state in the $+$ branch of
the superposition, and those in the subset $n$ \ follow the brain state in
the $-$ branch. The evolution of the minds is genuinely stochastic. This we
take to mean that before the minds actually diverge into the branches\ in
the state (\ref{eq:int1}), there is no determinate fact of the matter about
which branch each one of the minds will eventually follow. The membership of
a given mind in the subset $m$ (or $n$) becomes a fact at the same time that
the final state (\ref{eq:flip}) obtains. The standard quantum mechanical
probability is thus understood as the chance for each {\em single} mind to
end up in either the $m$-subset or the $n$-subset in the state (\ref{eq:flip}%
).

The motivation for assuming a multiplicity of minds here, rather than a
single mind that literally chooses one of the branches in the state (\ref
{eq:flip}), is this. First, in a single mind theory all the brain states of
the observer after a split are mindless, except for the one that is actually
tracked by the mind. This leads to the so-called {\em mindless hulk} problem
(Albert [1992], p. 130). Suppose, for example, that a second observer also
measures the $z$-component of spin when the state (\ref{eq:flip}) obtains,
and take $\alpha=\beta=1/\sqrt{2}$. Then the dynamical equations of motion
guarantee that the brain states of the two observers will be {\em correlated}
with certainty. But there is probability one-half that their minds will not
track the same branch of the state (see {\em ibid}, p. 130). Second, Bell's
theorem implies that in a single mind theory the {\em correlations} between
the minds of two observers on the two wings of a Bell-type experiment will
satisfy the quantum predictions only by allowing strong nonlocal dependence
between the trajectories of the minds (of the kind exhibited in usual hidden
variable theories; see Lockwood ([1996a])). The Albert and Loewer theory
avoids such strong nonlocality by associating with each brain state of an
observer after a split an infinity of minds, and by postulating a genuine
stochastic dynamics for the minds (see section \ref{chance} for more details
on this particular issue).

To sum up, we can characterize Albert and Loewer's interpretation as
follows. (i) There are no collapses, but the expansion of the global state (%
\ref{eq:flip}) in terms of the brain states $|\Phi _{i}\rangle $ and their 
{\em relative} states, e.\thinspace g.\ the pointer states $|\psi
_{i}\rangle $, is taken to describe our experience. (ii) There is a random
element built into the theory. The fact that the time evolution of the minds
is stochastic is depicted by the quantum mechanical probabilities. (iii) The
probability measure is {\em conditional} on a given measurement. (iv)
Individual minds (unlike the {\em proportion} of minds) do not supervene on
brain states. This means that an $m$-mind can be exchanged with an $n$-mind
in the superposition (\ref{eq:flip}) with no corresponding change in the
physics. In fact, the chance interpretation of the probability measure in
AL2 {\em implies} this failure of supervenience in Albert and Loewer's
version. This is the so-called {\em dualistic} aspect of this version
(Lockwood [1996a], Loewer [1996]).

Let us see how the Albert-Loewer approach bears on the relationship between
branching and relative frequencies. As is well known, this is a major
problem in the Everett picture where the number of branches resulting from a
given quantum measurement is not related to the quantum mechanical
probabilities. For example, in a measurement with two possible results the
quantum state will consist of two branches corresponding to the two results
of the measurement irrespective of the probabilities for each result. This
has the consequence that in a repeated measurement the relative frequencies
of an outcome (along a branch) will most likely mismatch the quantum
mechanical predictions. It then follows that the empirical success of
quantum mechanics, as observed by us, must be viewed as a miracle since most
of the Everett branches will not exhibit the right quantum mechanical
frequencies.

This problem can be solved if one postulates that the standard Born rule,
applied for each measurement, represents the probability of the branch. In
other words, one simply brings in the probability as an extra postulate in
addition to the branching. For example, consider a process where at $t_{1}$
a measurement with two possible results and probabilities $\frac{1}{3}$ and $%
\frac{2}{3}$ is performed. Then follows another measurement at $t_{2}$, with
three possible results with identical probabilities $\frac{1}{3}$, and so
on. A suitable law of large numbers can be proved for such a tree. In
particular, in a sequence of{\it \ }{\em identical} measurements, the
frequency on almost all branches will be close in value to the quantum
mechanical probability distribution on the set of measurement outcomes.
(e.\thinspace g.\ Everett ([1957]), DeWitt ([1970]), Hartle ([1968])). The
Albert-Loewer approach provides a simple explanation: Each individual mind
performs a (classical) random walk on the tree, with the probabilities
indicated on the branches. The fact that a typical mind perceives the
quantum mechanical frequencies simply follows from the theory of random
walks (or branching processes).

However,{\it \ }{\em quantum mechanics assigns, in advance, probabilities to
all possible measurement trees}. Given a quantum state of the system, we can
calculate in advance the probabilities of\ all possible sequences of
measurements. For example, insted of the measurement just considered, we can
perform at time $t_{1}$ a measurement with three possible outcomes whose
probabilities are $\frac{1}{5},\frac{3}{5},\frac{1}{5}$. At $t_{2}$ we do
not measure anything, and then at $t_{3}$ we perform a particular
measurement with two possible outcomes, and so forth. The probability for
each step in the sequence is known at the outset. Schr\"{o}dinger ([1935])
noticed that \ ''at no moment in time is there a collective distribution of
classical states which would be in agreement with the sum total of quantum
mechanical predictions''. This means that while each measurement sequence
can be seen as a classical random walk, there is no (non contextual)
classical probability distribution which assigns the correct probabilities
to all the branches of all possible trees simultaneously. This is a major
difference between the quantum concept of probability and the classical one.
One manifestation of this difference is the violation of Bell inequalities.
(Pitowsky [1994]).

We can see why the assumptions of Albert and Loewer are almost inevitable.
Suppose, contrary to AL2, that the trajectory of each mind is predetermined
before measurement. Now, consider Alice and Bob who participate in a typical
EPR experiment, Alice on the left and Bob on the right. In each run they can
each choose a direction along which to measure the spin. We now face the
task of choosing, {\em in advance}, the appropriate subsets for each
person's set of minds, corresponding to each possible result, in each
possible choice of directions. From Bell's theorem it follows that the only
way to do that, and obtain the right probabilities, is to violate Bell's
inequality. In the present context this means that the trajectories of some
Alice minds depend on Bob's choice of direction and vice versa. What we
have, in other words, is a non local hidden variable theory in disguise
(with the minds playing the role of hidden variables). Note that this formal
argument does not depend on any spatial characteristic of the minds
themselves, or lack thereof. This is part of the reason why Albert and
Loewer assume that the membership of a given mind in a given subset becomes
a fact only at the same time that the final state (e.g. state (\ref{eq:flip}%
)) obtains. That is, the random partition of the set of minds into the
appropriate subsets is conditional on which observable is actually measured.
They do not assume that there is a distribution on the set of minds that
explains the sum total of the quantum mechanical predictions.

\section{Probabilities in Lockwood's Version}

\begin{center}
\label{lock}
\end{center}

We now turn to analyzing the approach to probabilities in all version of the
many minds interpretation which assume complete supervenience of the mental
on the physical.\footnote{%
These versions include possibly the early formulations by Everett ([1957])
and DeWitt ([1970]), as well as later versions by Zeh ([1973]), Deutsch
([1985]), Zurek ([1993]), Donald ([1995], [1999]) Vaidman ([1998]), Saunders
([1998]), and also post-Everett approaches of consistent (decoherent)
histories, e.\,g.\ by Griffiths ([1984]), Gell-Mann and Hartle ([1990]).}
For convenience we shall focus on Lockwood's ([1996a,b]) quite explicit
approach to probabilities, but our analysis can be applied {\em mutatis
mutandis} to other versions. Lockwood aims explicitly at a picture in which
there is full supervenience of the minds on the brain states (or the
corresponding branches) and there is absolutely no stochastic behaviour of
the minds. Such versions assume:

\begin{description}
\item[SUP1]  The brain states $|\Phi _{\pm}\rangle $ corresponding to mental
states are associated at all times with a continuous infinity of nonphysical
entities called minds. The minds completely {\em supervene} on the brain
states $|\Phi _{\pm }\rangle$.
\end{description}

Lockwood's idea of supervenience of the minds on the $|\Phi _{\pm }\rangle $
means that he rejects the description of Albert and Loewer given by the
state (\ref{eq:flip}) and adopts instead the description given by (\ref
{eq:int2}) as a {\em complete} description of the physics {\em and} of the
minds. In other words, in Albert and Loewer's theory brain states are
indexed by subsets of minds, whereas in Lockwood's theory subsets of minds
are indexed by brain states.

To make sense of the quantum mechanical probabilities Lockwood defines a
probability measure over subsets of the minds as follows.

\begin{description}
\item[SUP2]  The standard Born rule defines a unique probability measure
over subsets of minds, such that for any measurement (involving a conscious
observer) the measure prescribes the {\em proportions} of minds following
each final branch of the superposition. For example, in the post-measurement
state (\ref{eq:int2}), the subset of minds following the $+$ branch is
assigned the measure $\left| \alpha \right| ^{2}$ and the subset of minds
following the $-$ branch the measure $\left| \beta \right| ^{2}$.
\end{description}

Thus, for Lockwood too, the quantum measure describes how many minds follow
each branch of the post-measurement state (such as state (\ref{eq:int2})).
Let us also note that Deutsch's ([1985]) probability measure, as well as
many other supporters of the supervenience view, is essentially the same,
except that it is sometimes defined over subsets of so-called {\em worlds}.

Lockwood's version differs from Albert and Loewer's with respect to the
interpretation of probability in that the evolution of the minds is {\em not}
genuinely random in the Albert-Loewer sense. However, in versions that
assume supervenience it is not at all clear what the dynamics of the minds
is. Nor is it clear how one could make the dynamics compatible with
supervenience. We consider below two possible interpretations.

1) Let $\Lambda $ be the set of minds of the observer. Minds supervene on
brain states. This means that with each brain state $\left| \Phi
\right\rangle ,$ which corresponds to a conscious perception of a
measurement outcome, Lockwood associates a subset $\ m(\left| \Phi
\right\rangle )\subset \Lambda $. Suppose that $\left| \Phi
_{+}\right\rangle $ is the brain state of an observer perceiving `spin up'
(for a quantum system and apparatus in a given state, on a given day, in a
given weather, and so on (including whatever it takes to specify the brain
state uniquely). Then the probability measure of the subset$\ m(\left| \Phi
_{+}\right\rangle )$ is $\left| \alpha \right| ^{2}$. Now, if we assume that
the association $\left| \Phi \right\rangle \rightarrow m(\left| \Phi
\right\rangle )$ of brain states with subsets of minds is fixed in advance
of any measurement, as suggested by supervenience (see {\bf SUP1}), we run
into a problem. We require, in fact, a choice of a probability measure on
the set of minds that will be in agreement with the totality of quantum
mechanical predictions. This implies non locality and contextuality in the
dynamics of the minds in the way that was explained at the end of the
previous section.\footnote{%
Notice that even if the minds' labels were to fix only probabilistically the
evolution of the minds, Bell's theorem would still apply, as long as the
postulation of a probability measure that explains all predictions is not
dropped.}

2) So perhaps one could endorse what Loewer ([1996]) calls the {\em %
instantaneous minds} view. (As far as we know Lockwood doesn't subscribe to
this view.) On this view it is assumed that minds do not persist in time in
the sense that there is no unique succession relation between any one of the
minds at an earlier time and any one of the minds at a later time. One could
think about this view as denying that a label can be attached to each of the
minds at an early time that would distinguish it from all but one of the
minds at a later time. As a result this view holds that there are no facts
of the matter concerning the evolution of a single mind between any two
times, and in particular between times during which the quantum state
evolves into superpositions of branches indexed by different mental states.

Consider for instance the time evolution taking the state (\ref{eq:int1}) at
time $t_{1}$ to the final state (\ref{eq:int2}) at time $t_{2}$. On the
instantaneous view all we can say about the behaviour of the minds is that
at $t_{1}$ the total set of the minds is associated with the state $%
|\Phi_{0}\rangle $, and at $t_{2}$ a subset of the minds with measure $%
\left|\alpha \right| ^{2}$ is associated with the final brain state $|\Phi
_{+}\rangle $ and a subset with measure $\left| \beta \right| ^{2}$ with the
brain state $|\Phi _{-}\rangle $. However, there is no fact of the matter as
to which mind in (\ref{eq:int1}) evolved into which branch in (\ref{eq:int2}%
) in the sense that there is no real mapping of the minds at $t_{1}$ to any
one of the two subsets in $t_{2}$ (see e.\,g.\ Lockwood ([1996a], p. 183),
but compare Lockwood ([1996b, pp. 458-9])).

It seems, however, that Lockwood would also want to maintain that $\Lambda $%
, the set of all minds, is itself time invariant. To put it differently,
minds are not created and destroyed through time (presumably, until the
person dies). Now, this is utterly inconsistent with the instantaneous minds
view. The first axiom of set theory, the axiom of equality, states that two
sets are equal when they have the same elements. If $\Lambda $ at $t_{1}$ is
identical to $\Lambda $ at $t_{2}$ there always exists a mapping, or a
succession relation, between the minds at the two times. Simply take the
identity mapping! Likewise, there is always a fact of the matter regarding
which element $\lambda \in \Lambda $ is an element of the subset $m(\left|
\Phi _{+}\right\rangle )$.

And so, if $\Lambda $ is time invariant, each mind in fact {\em is} labeled
through time just as Albert and Loewer insist. But then the question of
whether such minds evolve in a genuinely stochastic fashion or in a
deterministic fashion is still pertinent (see Butterfield [1996] for an
extended discussion of this point). In other words, on such a view one has
to provide a clear answer to the question whether the proposed dynamics of
the minds conforms to the Albert-Loewer stochastic type, in which case one
has to give up on supervenience. Alternatively, one could adopt the
deterministic version, in which case the evolution of the minds, in e.\,g.\
EPR situations, is sometimes determined by remote spacelike separated events.

But perhaps one would insist that $\Lambda $ is not time invariant, and at
each time $t$ there exists a different set of minds $\Lambda _{t}$. Minds
are born with every experiment, they briefly supervene on the brain state,
and then die like butterflies. What precisely such a theory {\em explains}
is not clear to us.

It is entirely possible that supporters of supervenience intend an
interpretation that is completely different from those that we have
discussed. Lockwood [(1996a, pp. 183-4)] seems to suggest that minds may be
like bosons having no individual identity. But this, really, seems to us a
conflation of the explanation with the problem. It is precisely the
strangeness of entangled states of the kind exhibited by identical bosons
which suggests theories like the many minds interpretation in the first
place. What would we achieve if the minds lived in a Hilbert space?

Nevertheless, it might be that some non-classical interpretation of
probability is compatible with the idea of supervenience in the Everett
tradition (see e.\thinspace g.\ Saunders [(1998]), Lockwood ([1996b])), and
would be fruitful in the sense that it includes enough of the content of our
usual, classically interpreted probabilistic assertions. However, as we have
just stressed, the Albert-Loewer {\em stochastic} and {\em non-supervening}
character of the minds, as stressed by Albert and Loewer, seems to be
necessary, if one wishes to have probabilities in the usual sense of the
word, while keeping the theory local in Bell's sense.

\section{Sets of Minds and Their Correlations}

\begin{center}
\label{chance}
\end{center}

We now turn to analyze more specifically the question of locality in the
Albert-Loewer version. Albert and Loewer ([1988]) and Albert ([1992, p.
132]) (see also Maudlin ([1994])) argue that their version is completely
local. On the other hand as we saw their chancy evolution of the minds is
designed to deliver the standard quantum mechanical predictions which, we
know, violate the Bell inequality. The stochastic evolution of the minds
which occurs only at or after an actual choice of measurement solves the
problem of Bell's nonlocality. However, in what follows we shall argue that
there is a {\em weaker }notion of locality that is violated even in the
Albert-Loewer version.

Let us start with Albert and Loewer's argument. Consider the singlet state
of the two particle system: 
\begin{equation}
\sqrt{1/2}\big(|+_{z}\rangle _{1}|-_{z}\rangle _{2}-|-_{z}\rangle
_{1}|+_{z}\rangle _{2}\big),  \label{eq:sin1}
\end{equation}
and suppose that observer 1 (Alice) measures some spin observable of
particle 1, and observer 2 (Bob) measures some (not necessarily the same)
spin observable of particle 2. The overall state after these two
measurements will be a superposition of branches (in general four branches).
In each branch each of the two measurements has some definite outcome. Now,
Albert and Loewer argue that no matter which observable gets measured by
Bob, the chances of Alice to see a $+$ result or a $-$ result are exactly
one-half. On this picture this means that one-half of the minds of Alice
will see an up result and one-half will see a down result {\em independently}
of the measurement of Bob. And the same goes for Bob.\footnote{%
And of course this fact does not depend on the time order of the two
measurements (or the reference frame in which we choose to describe the
measurements).} Moreover, Albert and Loewer stipulate that the evolution of
the minds of each observer (which recall is stochastic) is controlled by the
local {\em reduced} (physical) state of that observer {\em alone} (see
section \ref{al}). And as we have just argued this is sufficient to ensure
that the evolution of the minds will satisfy the frequencies predicted by
the quantum mechanical algorithm. According to Albert and Loewer, this is as
far as the many minds picture goes.

In particular, Albert and Loewer ({\em ibid}) (see also Albert ([1992], p.
132)) argue that there are just {\em no} matters of fact about the
correlations of the minds of Alice and Bob (nor about the outcomes of the
measurements). And if no such correlations obtain, we agree that the
Albert-Loewer picture is {\em ipso facto} local and Bell's theorem is simply
irrelevant at this stage. The correlations which Bell's theorem is about
will obtain only in a {\em local} way between the minds of one observer and
the {\em physical} reports of the other (see (B) below)). Once such a
correlation occurs the many minds picture gives the correct quantum
mechanical predictions in a completely local way.

We shall now argue, however, that there are {\em correlations} between
subsets of the minds of Alice and Bob. It is true that the chance
distribution of each of the observers' minds is independent of the
measurement on the other wing. For the reduced state of each of the
observers is independent of the interactions occurring on the other wing%
\footnote{%
This follows from the so-called no-signaling theorem in standard quantum
mechanics (see, e.\,g.\ Redhead ([1987]).}, and according to Albert and
Loewer's dynamical rule for the minds the evolution of the minds of each
observer is controlled solely by the reduced (local) state of that observer,
and is independent of the state of the other observer. With each experiment
performed by an observer the set of minds associated with the observer's
brain splits into many subsets, one for each possible outcome, and the
probability of each outcome is just the measure of the corresponding subset.

Take now an EPR setting in which Alice and Bob happen to measure the {\em %
same} observable. The overall state of the two particle system and the two
observers will evolve into a superposition of only two branches: 
\begin{equation}
\sqrt{1/2}\big(|+_{z}\rangle _{1}|-_{z}\rangle _{2}|\Phi _{+}\rangle
_{1}|\Phi _{-}\rangle _{2}-|-_{z}\rangle _{1}|+_{z}\rangle _{2}|\Phi
_{-}\rangle _{1}|\Phi _{+}\rangle _{2}\big).  \label{eq:sin2}
\end{equation}

(A) Consider a single run of the experiment. When Alice measures spin her
set of minds splits into two subsets of size one-half. Call them $A+$ and $%
A- $. Similarly for Bob, whose set of minds splits into $B+$ and $B-$ with
the same proportions. If we follow the track of a single Alice-mind we see
that it ends up in the subset $A-$ with probability one-half believing a $-$
result (and similarly for the minds in $A+, B+$ and $B-$). If the same
measurements are repeated $N$ times the minds in each of these subsets trace
paths connecting $N$ vertices on the binary tree of possible results.

(B) Consider for example Alice's minds in the subset $A-$ immediately after
her measurement. These minds may develop predictions about Bob's reports
concerning the result(s) of his measurement. In particular, an Alice-mind in 
$A-$ ($A+$) will predict with certainty that Bob will report a $+$ ($-$)
result. The same goes with respect to Bob's minds. According to quantum
mechanics these predictions will be confirmed with certainty. In order to
explain these predictions Albert and Loewer consider what happens to the
quantum state immediately after Alice and Bob communicate to each other
their results. The quantum mechanical equations of motion result in a state
with the same form as the state (\ref{eq:sin2}). In particular, in that
state Alice's brain state corresponding to the $-$ ($+$) result will be
perfectly correlated to Bob's {\em report} state corresponding to the $+$ ($%
- $) result. Therefore, when the final state obtains, the probability that
Alice's $A-$ ($A+$) minds will actually end up perceiving Bob as reporting a 
$+$ ($-$) result is equal to the usual quantum mechanical probability. That
is, in this case Alice's predictions of Bob's reports will be confirmed with
probability one. Thus, Albert and Loewer argue, there need be no facts about
correlations between the minds in $A-$ ($A+$) and the minds in $B+$ ($B-$)
in order to reproduce the quantum predictions (Albert [1992], p. 132)), for
it suffices that there is a correlation between minds and report-states.
Note that this applies both before and after the observers communicate to
each other their results.

So far this is uncontroversial. In the remainder of this paper, the weak
minds correlations we shall talk about can be understood as referring to the
above correlations between one observer's minds and another's reports.

(C) However, Albert and Loewer's conclusion that there are no correlations
between sets of minds is not entirely in line with other aspects of their
theory, namely their position on the mindless hulk problem (see section \ref
{al}). In particular, it is not clear why it was considered to be a problem
in the first place. The problem, recall, is that if each observer has only a
single mind, then there is probability of one-half (given the state (\ref
{eq:sin2})) that Alice's mind will follow the branch which is {\em not}
followed by Bob's mind.\footnote{%
Note that this is already a statement about correlations between the minds.}
For example, Alice's mind may perceive a $-$ result and so may Bob's mind.
When they meet, however, Alice's mind will with certainty perceive Bob as
reporting a $+$ result with certainty (this is the mindless hulk).

An analogous problem appears in Albert and Loewer's many minds theory.
According to Albert and Loewer there is no fact of the matter about whether
or not Bob's $+$ report, as perceived by Alice's $A-$ minds, corresponds to
Bob's $+$ minds. All we know is that Bob's report is associated with some
minds, but Albert and Loewer do not allow us to say that these minds are $+$
minds. Suppose there are only correlations between Alice's minds and Bob's
reports (and vice versa), as Albert and Loewer say, but not between Alice's
sets of minds and Bob's sets of minds. This means there will be a Bob $+$
mind who witnesses Alice reporting a $-$ result, while the latter report is
associated with a $+$ mind of Alice. This is analogous to the mindless hulk
problem. In the single mind case a mindless brain state is producing a
definite $-$ report , and in the many minds case a brain state associated
with a $+$ mind produces a $-$ report. We believe that if the first is a
problem, then so is the second.

In order to solve this problem we have to assume correlations between sets
of minds as given by the quantum mechanical predictions. We call these
correlations {\em weak minds-correlations} (weak nonlocality).\footnote{%
Note that in the versions by, e.\,g.\ Lockwood ([1996a]), Zurek ([1993]),
Vaidman ([1998]), Saunders ([1998]), Donald ([1999]), the weak
minds-correlations follow immediately from supervenience.} Notice that
without weak minds-correlations a single mind theory with the same {\em local%
} dynamics can reproduce the quantum predictions (and the minds-to-reports
correlations) just as well.\footnote{%
The argument for this is the same as the one given in (B) above in the case
of the many minds theory.}

Thus in the many minds picture the (weak) minds-correlations are necessary
in order to provide a genuine solution to the mindless hulk problem. These
correlations allow us to say, in the EPR case above for example, that the
path of an Alice mind in $A-$ ($A+$) is the mirror image of the path of a
Bob mind in $B+$ ($B-$). It is in fact an advantage of Albert and Loewer's
theory that introducing such correlations doesn't violate locality in Bell's
sense. This is possible because the multiplicity of the minds, as it were,
compensates for the local dynamics. Bell's theorem requires the existence of
a single probability measure defined over all possible (actual and
counterfactual) measurements. This is denied by Albert and Loewer's local
dynamics of the minds. In this way the derivation of a Bell inequality is
formally blocked.

\section{Many Minds and GHZ}

\begin{center}
\label{ghz}
\end{center}

An even more revealing case of weak nonlocal correlations between the minds
in Albert and Loewer's theory is that of Greenberger, Horne and Zeilinger
(GHZ) ([1989]) (See also Mermin ([1990])). In this case we will show that
the quantum mechanical correlations and algebraic relations between the
observables entail nonlocal constraints on the distribution of the minds
into subsets corresponding to the results of measurements. Moreover, since
in Albert and Loewer's theory the individual minds retain their identity
throughout time, these constraints have an implication even for {\em %
individual} minds.

Consider the GHZ-state 
\begin{equation}
\sqrt{1/2}\big(|+_{z}\rangle _{1}|+_{z}\rangle _{2}|+_{z}\rangle
_{3}-|-_{z}\rangle _{1}|-_{z}\rangle _{2}|-_{z}\rangle _{3}\big),
\label{eq:ghz}
\end{equation}
where the kets $|\pm _{z}\rangle _{i}$ ($i=1,2,3$) denote the $z$-spin state
of particle $i$. Suppose that the three particles are located in space-like
separated regions. In this state standard quantum mechanics predicts a $+$
or $-$ result of the local measurements of the $z$-spin of each particle
with probability one-half (with collapse onto the corresponding branch). And
likewise for measurements of the $x$- and $y$-spins of each particle.

However, for this specific state standard quantum mechanics also predicts
certain correlations between the results of the three observers, in
particular the correlations described by the algebraic relations 
\begin{equation}
\begin{array}{lcl}
{\rm (XXX)}\;\;\;\; \sigma _{x}^{1}\sigma _{x}^{2}\sigma _{x}^{3} & = & -1 \\%
[0.5ex] 
{\rm (XYY)}\;\;\;\; \sigma _{x}^{1}\sigma _{y}^{2}\sigma _{y}^{3} & = & 1 \\%
[0.5ex] 
{\rm (YXY)}\;\;\;\; \sigma _{y}^{1}\sigma _{x}^{2}\sigma _{y}^{3} & = & 1 \\%
[0.5ex] 
{\rm (YYX)}\;\;\;\; \sigma _{y}^{1}\sigma _{y}^{2}\sigma _{x}^{3} & = & 1
\end{array}
\label{eq:corr}
\end{equation}
It is easy to see that the state (\ref{eq:ghz}) is an eigenstate of the
above four observables with eigenvalues indicated on the right. Assuming
that the local observables $\sigma _{x}^{i},$ $\sigma _{y}^{i}$ take on
definite values that are fixed {\em locally},{\em \ }and that the values
satisfy the correlations (\ref{eq:corr}), we can easily derive the
contradiction 
\begin{equation}
-1=\sigma _{x}^{1}\sigma _{x}^{2}\sigma _{x}^{3}=\sigma _{x}^{1}\sigma
_{y}^{2}\sigma _{y}^{3}=1.  \label{eq:op}
\end{equation}
This is the GHZ simplification of Bell's theorem.

In standard quantum mechanics (with a collapse postulate) this contradiction
is avoided, since the local observables $\sigma_{x}^{i},\sigma _{y}^{i}$ are
assigned no definite individual values in the initial state (\ref{eq:ghz}),
and the collapse itself is {\em nonlocal} in the following sense. The result
of the (local) measurements $\sigma _{x}^{i}$ or $\sigma _{y}^{i}$ on each
wing of the experiment depends on which observables get measured on the two
other wings and on their outcomes. The correlations between the results of
the measurements on the three wings satisfy exactly the correlations given
by (\ref{eq:corr}).

But now what happens in Albert and Loewer's many minds theory in which there
is no collapse of the state? If the three observers measure spin along the $%
{\rm XXX}$-directions, the uncollapsed quantum state, when these
measurements are over, must be: 
\begin{eqnarray}
\vert \Psi\rangle &=& 1/2 (|+_{x}\rangle _{1}|-_{x}\rangle
_{2}|+_{x}\rangle_{3}|\Phi _{+}\rangle _{1}|\Phi _{-}\rangle _{2}|\Phi
_{+}\rangle _{3}+  \label{eq:x} \\[2ex]
&& +|-_{x}\rangle_{1}|+_{x}\rangle _{2}|+_{x}\rangle _{3} |\Phi
_{-}\rangle_{1}|\Phi _{+}\rangle _{2}|\Phi _{+}\rangle _{3}+  \nonumber \\%
[2ex]
&& +|+_{x}\rangle _{1}|+_{x}\rangle _{2}|-_{x}\rangle _{3}|\Phi
_{+}\rangle_{1}|\Phi _{+}\rangle _{2}|\Phi _{-}\rangle _{3}+  \nonumber \\%
[2ex]
&& +|-_{x}\rangle _{1}|-_{x}\rangle _{2}|-_{x}\rangle _{3}|\Phi
_{-}\rangle_{1}|\Phi _{-}\rangle _{2}|\Phi _{-}\rangle _{3})  \nonumber
\end{eqnarray}
(written in the $x$-spin bases).

In this state the marginal probabilities for a $+$ or $-$ result of the
local measurements of each observer is one-half. In the Albert and Loewer
theory this means that for each observer the proportions of the minds
perceiving a $+$ or $-$ result are one-half. But because of the constraint $%
{\rm XXX}$ in (\ref{eq:corr}) the state (\ref{eq:x}) consists of only four
branches in each of which there is a strict anti-correlation between the
sign of the result perceived by, say the subset of minds of observer 1, and
the sign of the product of the results perceived by subsets of minds of
observers 2 and 3. These anti-correlations are brought about since the
quantum mechanical amplitudes of the branches with strict correlations in
the state (\ref{eq:x}) is zero. Now as we have argued in section \ref{chance}
the anti-correlations in the state (\ref{eq:x}) mean that the {\em subsets
of minds} of the three observers that correspond to these branches are
weakly correlated (in the sense of (C) in section \ref{chance}). A similar
analysis applies for the alternative measurements along the directions ${\rm %
XYY}$, ${\rm YXY}$ and ${\rm YYX}$, but in these latter cases quantum
mechanics predicts strict correlations between the (sign of the) results
perceived by subsets of minds of any given observer and the sign of the
product of the results perceived by subsets of minds of the other two
observers (see equation (\ref{eq:corr})). Call the four possible
measurements ${\rm XXX}$, ${\rm XYY}$, ${\rm YXY}$ and ${\rm YYX}$ {\em %
scenarios} and number them from $1$ to $4$ respectively.

Consider the implications of the GHZ analysis for the Albert and Loewer
theory. Given the quantum mechanical correlations described by (\ref{eq:corr}%
), we obtain that in each scenario the set of triples of minds $M_{A}\times
M_{B}\times M_{C}$ (where $M_{A}$ is the set of Alice's minds, etc.) is
partitioned into four disjoint subsets. Thus, in the first scenario ${\rm XXX%
}$ we can write

\begin{eqnarray}
M_{A}\times M_{B}\times M_{C} &=& M_{A}^{+}(1)\times M_{B}^{+}(1)\times
M_{C}^{-}(1) \cup \\
&&\cup M_{A}^{+}(1)\times M_{B}^{-}(1)\times M_{C}^{+}(1)\cup  \nonumber \\
&& \cup M_{A}^{-}(1)\times M_{B}^{+}(1)\times M_{C}^{+}(1)\cup  \nonumber \\
&& \cup M_{A}^{-}(1)\times M_{B}^{-}(1)\times M_{C}^{-}(1).  \nonumber
\end{eqnarray}

Here, $M_{A}^{+}(1)$ stands for the subset of Alice's minds that observe $+$
in the first scenario. In the other three scenarios ${\rm XYY}$, ${\rm YXY}$
and ${\rm YYX}$ we have

\begin{eqnarray}
M_{A}\times M_{B}\times M_{C} &=&M_{A}^{+}(j)\times M_{B}^{-}(j)\times
M_{C}^{-}(j)\cup \\
&&\cup M_{A}^{-}(j)\times M_{B}^{+}(j)\times M_{C}^{-}(j)\cup  \nonumber \\
&&\cup M_{A}^{-}(j)\times M_{B}^{-}(j)\times M_{C}^{+}(j)\cup  \nonumber \\
&&\cup M_{A}^{+}(j)\times M_{B}^{+}(j)\times M_{C}^{+}(j),  \nonumber
\end{eqnarray}
where $j=2,3,4$ index the scenario.

We stress that this fact, namely that the set of triples of the minds is
partitioned into four disjoint subsets in each scenario, is completely {\em %
independent} of the question which scenario (if any) will be eventually
performed. We may assume, for example, that this question is settled {\em %
deterministically} by the initial conditions. This assumption is perfectly
consistent with Albert and Loewer's theory.

To see how the weak non-locality is manifested here consider, for example,
the set

\begin{eqnarray}
&&M_{A}^{-}(1)\times M_{B}^{+}(1)\times M_{C}^{+}(1)\cap M_{A}^{-}(2)\times
M_{B}^{+}(2)\times M_{C}^{-}(2)\cap \smallskip  \label{eq:intersec} \\
&&\cap M_{A}^{-}(3)\times M_{B}^{-}(3)\times M_{C}^{+}(3)\cap
M_{A}^{+}(4)\times M_{B}^{+}(4)\times M_{C}^{+}(4).  \nonumber
\end{eqnarray}
It consists of an intersection of four partition elements, one from each
scenario. If a triple of minds belongs to this set then Bob's mind perceives 
$+$ in the $x$-direction in scenario $1$ and $-$ in the $x$-direction in
scenario $3$, while Carol's mind perceives $-$ in the $y$-direction in
scenario $2$ and $+$ in the $y$-direction in scenario $3$. By GHZ, this is
true for any set like (\ref{eq:intersec}), an intersection of four partition
elements, one from each scenario{\em .} For each triple in such a set at
least one of the observers flips signs in the same local measurement, while
changing the scenario{\em .} In other words, either Alice's mind is flipping
when changing from scenario $1$ to $2$, or from scenario $3$ to $4$; or
Bob's mind is flipping when changing from $1$ to $3$, or from $2$ to $4$; or
Carol's mind is flipping when changing from $1$ to $4$, or from $2$ to $3$.

We can estimate the size of intersections like (\ref{eq:intersec}). There
are $4^{4}=256$ such sets, and between them they cover all the logical
possibilities. Therefore, at least one of those sets has probability $\geq 
\frac{1}{256}$. This type of information is available in Albert and Loewer's
theory, while completely absent from quantum mechanics. In this sense their
interpretation is a hidden variable theory of sorts.

We stress that the above argument cannot be circumvented by merely pointing
out that the evolution of the minds in the Albert and Loewer theory is
stochastic. As can be seen in the previous paragraph the fact that any given
mind has probability of $1/2$ to perceive a $+$ or a $-$ result in each
scenario is already taken into account in our argument. The weak nonlocal
correlations between the minds is not only a statistical constraint on the 
{\em proportions} of minds. Rather, it is a constraint on the correlations
between {\em individual minds across scenarios.}

However, since the dynamics of the minds in the Albert and Loewer theory is
genuinely stochastic, it is contingent which of the individual minds of each
observer flip their sign between different scenarios. In other words, {\em %
our argument does not involve counterfactual reasoning about individual minds%
}. Hence, our use of the term `scenario', and not `possible world'. The
latter would appropriately correspond to a particular assignment of a value
to each individual mind in a particular scenario. What changes from one
scenario to another are the {\em sets of minds}, and each scenario
corresponds, therefore, to a multitude of possible worlds. The weak sense of
nonlocality obtained in this way is hidden behind the stochastic evolution
of the minds, and eventually behind the stochastic results we perceive.

Finally, since the weak form of these nonlocal correlations is a feature of
the {\em uncollapsed} quantum state, such as (\ref{eq:x}), the evolution of
the minds need not exhibit, not even in principle, a dependence on a
reference frame. In this sense a theory with such correlations might be
written in a genuine relativistic form. But this is another issue.

\begin{center}
Acknowledgements
\end{center}

\noindent We thank David Albert, Jeremy Butterfield, Matthew Donald, Barry
Loewer, Simon Saunders and David Wallace for their comments. Itamar Pitowsky
thanks the Israel Science Foundation for a grant number 787/99, which
supports this research.

\begin{flushright}
\noindent Meir Hemmo\\
Department of Philosophy\\
University of Haifa\\
Haifa 31905, Israel\\
email: meir@research.haifa.ac.il
\end{flushright}

\begin{flushright}
\noindent Itamar Pitowsky\\
Department of Philosophy\\
The Hebrew University\\
Jerusalem 91905, Israel\\ 
email: itamarp@vms.huji.ac.il
\end{flushright}

\begin{center}
References
\end{center}

\noindent Albert, D. [1992]: {\em Quantum Mechanics and Experience},
Cambridge, Mass.: Harvard University Press.

\noindent Albert, D. and Loewer, B. [1988]: `Interpreting the Many Worlds
Interpretation', {\em Synthese}, {\bf 77}, \mbox{pp. 195-213}.

\noindent Butterfield, J. [1996]: `Whither the Minds?', {\em British Journal
for the Philosophy of Science}, {\bf 47}, \mbox{pp. 200-21}.

\noindent Deutsch, D. [1985]: `Quantum Theory as a Universal Physical
Theory', {\em International Journal of Theoretical Physics} {\bf 24}, %
\mbox{pp. 1-41}.

\noindent DeWitt, B. [1970]: `Quantum Mechanics and Reality', {\em Physics
Today}, {\bf 23}, \mbox{pp. 30-5}

\noindent Donald, M. J. [1995]: `A Mathematical Characterization of the
Physical \linebreak Structure of Observers', {\em Foundations of Physics}, 
{\bf 25}, \mbox{pp.529-71}.

\noindent Donald, M. J. [1999]: `Progress in a Many Minds Interpretation of
Quantum Theory', {\em Los Alamos E-Print Archive},
http://www.arxiv.org/abs/quant-ph/\linebreak 9904001.

\noindent Everett, H., III [1957]: `Relative State Formulation of Quantum
Mechanics', {\em Reviews of Modern Physics}, {\bf 29}, \mbox{pp. 454-62}.

\noindent Gell-Mann, M. and Hartle. J. [1990]: `Quantum Mechanics in the
Light of Quantum Cosmology', in W. Zurek ({\em ed.}), {\em Complexity,
Entropy and the Physics of Information, SFI Studies in the Physics of
Complexity} (Vol. VIII), Redwood City, CA: Addison-Wesley, \mbox{pp.425-58}.

\noindent Greenberger, D. M., Horne, M. and Zeilinger, A. [1989]: `Going
Beyond Bell's Theorem', in M. Kafatos ({\em ed.}) {\em Bell's Theorem,
Quantum Theory and Conceptions of the Universe}, Dordrecht: Kluwer Academic, %
\mbox{pp. 69-72}.

\noindent Griffiths, R. [1984]: `Consistent Histories and the Interpretation
of Quantum Mechanics', {\em Journal of Statistical Physics}, {\bf 36}, %
\mbox{pp. 219-72}.

\noindent Hartle, J. [1968]: `Quantum Mechanics of Individual Systems', {\em %
American Journal of Physics}, {\bf 36}, \mbox{pp. 704-12}.

\noindent Lockwood, M. [1996a]: `Many Minds Interpretations of Quantum
Mechanics', {\em British Journal for the Philosophy of Science}, {\bf 47}, %
\mbox{pp.159-88}

\noindent Lockwood, M. [1996b]: `Many Minds Interpretations of Quantum
Mechanics: Replies to Replies', {\em British Journal for the Philosophy of
Science}, {\bf 47}, \mbox{pp.445-61}

\noindent Loewer, B. [1996]: `Comment on Lockwood', {\em British Journal for
the Philosophy of Science}, {\bf 47}, \mbox{pp. 229-32}.

\noindent Maudlin, T. [1994]: {\em Quantum Nonlocality and Relativity},
Oxford: Blackwell.

\noindent Mermin, D. [1990]: `Simple Unified Form of the Major No-Hidden
Variables Theorems', {\em Physical Review Letters}, {\bf 65}, 
\mbox{pp.
3373-6}.

\noindent Papineau, D. [1996]: `Many Minds Are no Worse Than One', {\em %
British Journal for the Philosophy of Science}, {\bf 47}, \mbox{pp. 233-41}.

\noindent Pitowsky, I. [1994]: `George Boole's `Conditions of Possible
Experience' and the Quantum Puzzle', {\em British Journal for the Philosophy
of Science}, {\bf 45}, \mbox{pp. 95-125}.

\noindent Redhead, M. [1987]: {\em Incompleteness, Nonlocality and Realism},
Oxford:\linebreak Clarendon Press.

\noindent Saunders, S. [1998]: `Time, Quantum Mechanics and Probability', 
{\em Synthese}, {\bf 114}, \mbox{pp. 373-404}.

\noindent Schr\"{o}dinger, E. [1935]: `Die Gegenvartige Situation in der
Quantenmechanik', in A. Dick, G. Kerber and W. Kerber ({\em eds.}), {\em %
Erwin Schr\"{o}dinger's Collected Papers} (Vol. 3), Vienna: Austrian Academy
of Sciences, \mbox{pp. 484-501}.

\noindent Vaidman, L. [1998]: `On Schizophrenic Experiences of the Neutron
or \linebreak Why We Should Believe in the Many-Worlds Interpretation of
Quantum Mechanics', {\em International Studies in the Philosophy of Science}%
, {\bf 12}, \mbox{pp.245-61}.

\noindent Zeh, H. [1973]: `Toward a Quantum Theory of Observation', {\em %
Foundations \linebreak of Physics}, {\bf 3}, \mbox{pp. 109-16}.

\noindent Zurek, W. [1993]: `Preferred States, Predictability, Classicality,
and the Environment-Induced Decoherence', {\em Progress in Theoretical
Physics}, {\bf 89}, \linebreak \mbox{pp. 281-312}.

\end{document}